# Deep Learning-Based Approach for Improving Relational Aggregated Search

Sara Saad Soliman[1], Ahmed Younes[1,2], Islam Elkabani [1,3] and Ashraf Elsayed [1,2]
[1] Department of Mathematics and Computer Science, Faculty of Science, Alexandria University, Alexandria 21511, Egypt
[2] Faculty of Computer Science and Engineering, Alamein International University, Al Learning-Alamein 51718, Egypt
[3] School of Information Technology, University of Cincinnati, Ohio, USA

## Abstract

Due to an information explosion on the internet, there is a need for the development of aggregated search systems that can boost the retrieval and management of content in various formats. To further improve the clustering of Arabic text data in aggregated search environments, this research investigates the application of advanced natural language processing techniques, namely stacked autoencoders and AraBERT embeddings. By transcending the limitations of traditional search engines—which are imprecise, not contextually relevant, and not personalized—we offer more enriched, context-aware characterizations of search results, so we used a K-means clustering algorithm to discover distinctive features and relationships in these results, we then used our approach on different Arabic queries to evaluate its effectiveness. Our model illustrates that using stacked autoencoders in representation learning suits clustering tasks and can significantly improve clustering search results. It also demonstrates improved accuracy and relevance of search results.

## 1  Introduction

The growing availability of information on the Internet has made aggregated search an essential strategy for enhancing the retrieval and management of various types of content [1]. Aggregate search collects and combines information from multiple sources, allowing users to access relevant content from different domains. This approach improves the user experience by providing appropriate information and tackling the challenges posed by increasing data diversity and volume. Users can acquire more information that meets their needs because aggregated search systems integrate data formats such as text, images, and videos to deliver precise and comprehensive results [2].

Key to the success of embedding models in natural language [3] is their translation of unstructured text into something we can feed into machine learning. Of all the models, AraBERT is the most widely used language representation model that is pre-trained on Arabic texts. AraBERT utilizes contextual embeddings that build semantic relationships within the text and capitalize on the nuances of the Arabic language—all based on transformer architecture.

This approach offers a more comprehensive understanding of Arabic and is beneficial for sentiment analysis, text classification, information retrieval, and more [4].

Like a neural network architecture called the stacked autoencoder model, which is utilized for feature extraction and dimensionality reduction, AraBERT performance can be improved with it. Levels on levels; each level output vector is the input to another autoencoder, retrieving an increasingly sophisticated compression. This allows the model to identify better complex patterns in the high dimensionality of the input data, resulting in more significant feature extraction. Stacked autoencoders can capture useful feature embeddings that significantly improve the performance of tasks from clustering to classification [5].

Clustering is the basis of group analysis and learning. It goes through the objects in a structured manner and clusters them according to their similarities. Clustering classifies data into different groups so that items in the same group are more like each other than items in other groups. This type of learning that does not require supervision is commonplace in many areas, including advertising, biology, social science, and image processing [6].

There are different clustering algorithms developed to be used for a particular problem domain, such as [7]:

1. **Partitioning Techniques:** Including K-Means, the algorithms split the data set into n clusters, n of which is already predefined. They repeatedly allocate data points to the clusters and reposition the centroids to minimize the difference between the points within the cluster.
2. **Hierarchical Techniques:** Cluster structures are organized in a tree-like format that allows users to view the data at multiple detail levels. This category includes both agglomerative and divisive approaches.
3. **Density-Based Techniques:** These include DBSCAN and its counterparts, which form clusters around regions where the density of the data points is significantly higher than in other areas. These algorithms are especially adept at identifying clusters of random shapes while minimizing noise.
4. **Model-Based Methods**: These methods, such as Gaussian mixture models (GMM), assume that the data are derived from a set of probability distributions.

In this work, we focus on extracting distinct features associated with each search vertical result between vertical search results for the Arabic language and identifying the diverse relationships between these vertical results. To address this challenge, we introduce an approach that utilizes aggregated search for the Arabic language with the AraBERT Embedding technique to generate more efficient representations for clustering the various vertical search results. We used the K-means algorithm, an efficient and easy-to-implement model commonly used to find underlying patterns in data when no class label is provided.

The remainder of this paper is organized as follows. The next section reviews the related work on Aggregated Search (AS). Section 3 describes the main problem addressed by this paper. We present our proposed method in Section 4. In Section 5, we display our experiment. Finally, we present the conclusion and discuss future research.

## 2  Related Work

An aggregate search combines search results from different search areas and displays them as a single result in a unified user interface.

This approach improves the search experience by providing a comprehensive, unified view of related information. It addresses the limitations of traditional search engines when dealing with different types of data and search tasks. Many studies and research papers have been conducted and published on this topic.

[8] Using semantics, they proposed a relational aggregated search approach for the disjoint multimedia search results. The disconnected multimedia vertical samples are combined using a relational aggregation approach, which then merges them and re-ranks them into a linearly ordered list according to how relevant they are to the user's query. This improves multimedia searches and preserves the relationships between multimedia content.

[9] enhanced vertical selection in aggregate search shows a technique for predicting the user's vertical intent. The method improves the query representation with rich semantic information by using a convolutional neural network (CNN) model with the Doc2vec algorithm to create query embedding vectors. Experiment results show how well the system provides accurate predictions while saving users time and effort when browsing. Additionally, it provides accurate predictions on new, unseen data.

This method [10] compares learning ranking algorithms and interleaving techniques in relational aggregate search (RAS). The emphasis is on presenting information to users and which method is most effective for ranking tasks. According to experimental results, LambdaMART outperforms other algorithms, whereas VI-TDA exceeds TDA in satisfying users' information needs.

In [11], an architecture for aggregated web search results was proposed to explore and discover multimedia documents. Semantic analysis and nonlinear navigation patterns allow users to explore and navigate multimedia content effectively. Empirical evaluations indicate that this approach achieves high accuracy in aggregated search results and provides a valuable foundation for investigating these results in multimedia contexts.

Another work [12] explores how users discover embedded multimedia content online. It analyzes users' challenges searching for multimedia information and suggests new tools to improve the discovery process. The research evaluates the performance of current search engines compared to the best discovery search systems and provides recommendations for strengthening typical web search engines.

In [13], AMED conducts deep semantic analysis, clustering, and summarization on SERP segments to enable multimedia document-based traversal. To engage users with the search area, they connected the multimedia documents to a non-linear graph based on similarity measures and presented them on interactive search user interfaces (SUIs). The experimental evaluation of the proposed data model surpasses clustering techniques, achieving an accuracy of up to 99%.

Another technique in [14] is the MIRRE method, which focuses on non-linear and multimodal browsing of combined multimedia search results. The technique leverages multimodal similarities and semantic relations to construct a result space with an SUI design.

The work in [15] addresses the issue of extracting valuable knowledge from unstructured text data. It proposes a novel deep text clustering method that combines stacked autoencoders with k-means clustering and adopts BERTopic modeling for cluster analysis.

# 3 Description of the research problem

Traditional search methods come with several drawbacks and difficulties. Some of the major problems in conventional search include:

Traditional search engines have been widely used to retrieve information from the vast web content. However, traditional research faces numerous limitations and challenges. One of the main concerns is the balance between recall and precision. While standard search engines strive to deliver accurate results, they often sacrifice memorability, leading to overlooking crucial information for specific searches. Users who depend on thorough and precise search results may find this frustrating. Another limitation of traditional research is the lack of contextual awareness.

Traditional search engines treat queries as isolated units, ignoring the overall context and user intentions behind them. As a result, search results may not effectively meet users' specific information needs or provide relevant contextual details. This limitation reduces the effectiveness and accuracy of search results, especially for ambiguous or complex queries. Moreover, traditional search engines rely heavily on keyword matching, which can create gaps in language and semantics. Natural language is intricate, and keyword matching fails to capture its subtleties or assess the semantic meanings of documents and queries. Consequently, this can lead to inaccurate or irrelevant search results, as the search engine cannot fully grasp the user's intentions or the query context. Web content has increased exponentially, leading to information overload. Traditional search engines face challenges in managing the vast amount of information, making it hard for people to find what they need. Users are overwhelmed with countless results and sifting through irrelevant content to find valuable information that helps them achieve their goals is daunting.

Lastly, traditional search engines lack personalization. They often provide identical search results to all users, regardless of their interests, preferences, or previous interactions. This one-size-fits-all approach overlooks the importance of tailoring the search experience for each user, leading to generic results that fail to meet their specific needs or context.

An aggregated search is proposed as an effective solution to address these limitations and challenges. This approach aims to go beyond the constraints of conventional search by consolidating results from various search engines and sources. By integrating multiple sources and utilizing advanced techniques such as natural language processing and machine learning, aggregated search can improve contextual understanding, bridge language and semantic gaps, and deliver more accurate and relevant search results.

Integrating AraBERT embeddings with stacked autoencoders presents a promising approach for optimizing aggregated search systems focusing on Arabic content. This combination enhances feature extraction by providing dense contextual text representations, which makes clustering algorithms more efficient. By utilizing this integrated strategy, we aim to improve information retrieval in aggregated search systems, allowing users to gain insights from various unstructured Arabic data sources. This paper describes how aggregated search, AraBERT embeddings, and stacked autoencoders can be combined to enhance Arabic textual information retrieval and clustering, ultimately delivering more effective and user-centered search experiences.

## 4 Proposed method

This section provides a detailed description of the proposed process flow. The proposed approach can be divided into four main stages, as shown in Fig. 1: Preparing the Dataset, Results Embedding (AraBERT), Feature Extraction (Stacked autoencoder), and Clustering Algorithm (K-means clustering).

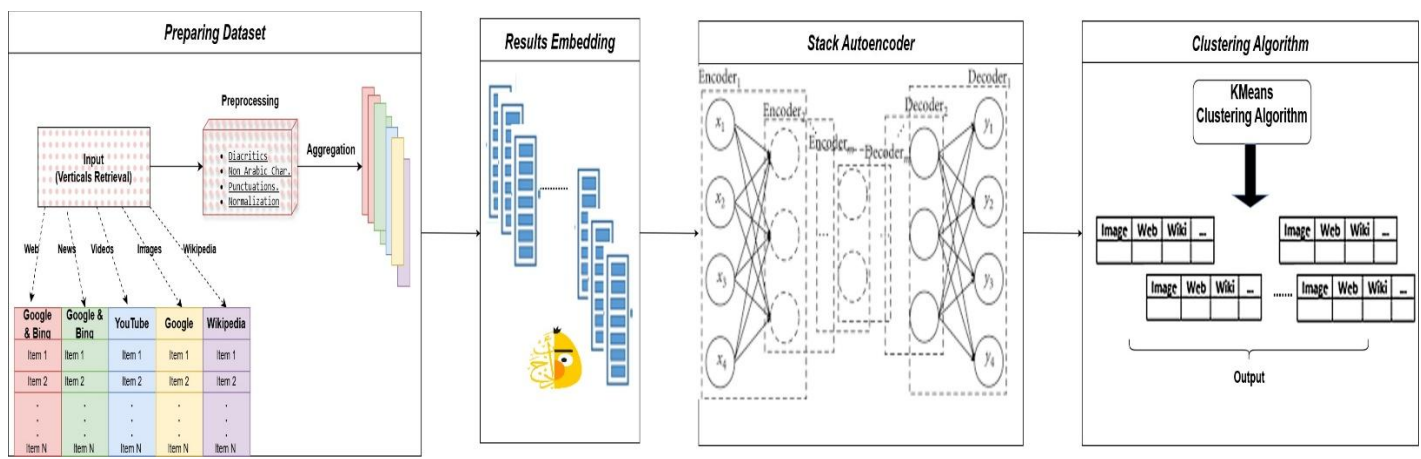

*Fig. 1: Proposed Method Architecture*

- **Preparing Dataset:** The first task in this step is to represent the data, as the results from various sources are heterogeneous; for example, they can include text data, multimedia content in the form of images and videos, or both. The task is to combine these various forms of content into a single consistent format. For this, we represented the data textually by extracting textual descriptions from each multimedia result and normalizing these representations, where the web can be defined as (title, link, and snippets), the image can be represented as (title, and link), and videos can be represented as (title, link, and description). Then, we applied text preprocessing techniques to clean up the data. We preprocessed the documents by removing diacritics and non-Arabic characters, including numbers and URL links, and removed Punctation Further and Normalization. Stop words step is not used because it is crucial to the meaning of the documents when using the AraBERT model. This consistent format is fed into the Stacked Autoencoder.

- **Results Embedding Using AraBERT:** The generated textual representation of the vertical results serves as input for the Stacked Autoencoder (SAE). However, for neural networks, inputs must be fixed-length feature vectors before being processed by the network. In the Vector Space Model (VSM), a document is represented as a vector, with each term functioning as a feature accompanied by assigned weight values. As a result, a single document can yield a substantial number of features. The dimensionality of the feature space and the occurrence of uninformative features can significantly affect the performance of the clustering algorithm. In this phase, we utilized the pre-trained AraBERT as our embedding algorithm to generate contextualized sentence embeddings. This model extracts features in the form of word and sentence embedding vectors from the data, producing fixed-length tensors (which comprise 12 encoder blocks, 768 hidden dimensions, 12 attention heads, a maximum sequence length of 512, and a total of approximately 110M parameters) [21]. In our study, we evaluated the effectiveness of the AraBERT model in the results embedding stage.

- **Feature extraction via stacked Autoencoders:** Extracting features is essential for clustering because it influences the quality and relevance of the data used to group similar items. Effective extraction transforms raw data into structured formats, enabling clustering algorithms to identify patterns and relationships swiftly. This allows the algorithm to create accurate and interpretable clusters. In this research, we utilized an efficient method for feature extraction known as "Stacked Autoencoder models." These neural networks compress data into lower-dimensional representations and then reconstruct the original input, which helps eliminate noise and capture data structures. This aided our research in extracting valuable features while removing irrelevant details that improve clustering performance. The autoencoder model comprises two primary components: the encoder and the decoder. These components work together but perform different functions to represent and reconstruct data. The encoder compresses the input data into a smaller format while the decoder rebuilds the data, retaining the essential elements. This unique structure makes autoencoders a powerful tool for feature extraction and various applications in machine learning. The activation function used in this study is the rectifier or rectified linear unit (ReLU), which introduces non-linearity and mitigates vanishing gradients in deep learning models. These models consist of multiple layers of stacked autoencoders, where each hidden layer's output connects to the Next's input. Each layer is trained individually in a stacked autoencoder, one at a time. The parameters used in each layer are shown in **Table 1.**

- **K-Means Clustering**: Text clustering is a widely used text mining method in which documents are grouped based on their similarities. The primary objective is to merge similar documents while differentiating them from those in distinct groups. In the final step of the proposed model, we input the features obtained from the previous step into the K-Means clustering algorithm. K-Means is one of the most popular and straightforward clustering algorithms. Since it does not use labeled data, it functions unsupervised; therefore, no individual document is assigned to any specific class or group. Text clustering is a field where K-Means is exceptionally popular.

# 5 Experimental results

In this section, we outline the experimental setup. The experiments used Python code on the TensorFlow framework, leveraging an i7 core processor. The development environment was Google Colab with GPU support, operating on a 64-bit Windows 10 system with 16GB of RAM.

## 5.1 Data set

- This section details the data utilized in this paper. To evaluate the performance of the proposed clustering method, we conducted experiments across seven verticals constructed using free APIs from Google (web search, images, news), Bing (web search, news), YouTube, and Wikipedia. We considered all relevant and irrelevant results for various queries per vertical search. In this case, we deal with documents retrieved from search engines for given different queries such as ("تكنولوجيا المعلومات", "الرياضه", "التعليم"), and in this case, we will try our work on four different queries ("التعليم", "الرياضه", "تكنولوجيا المعلومات"), ("تكنولوجيا", "الرياضه", "التعليم", "المعلومات"). Queries were chosen to test our work with data that is partly close in terms of the topic ("التعليم", "تكنولوجيا المعلومات"), ("التعليم", "الرياضه"). and data that is completely different in terms of subject ("الرياضه", "تكنولوجيا المعلومات"), ("الرياضه", "التعليم", "تكنولوجيا المعلومات"). with a different number of clusters Consequently, we opted to present the data in textual form by utilizing the textual description of each multimedia content result and normalizing the representation of the various outcomes before inputting them into the Deep Autoencoder. i.e., we have four queries where:
  - Q1: ("التعليم", "الرياضه"),
  - Q2: ("الرياضه", "تكنولوجيا المعلومات"),
  - Q3: ("التعليم", "تكنولوجيا المعلومات"),
  - Q4: ("التعليم", "الرياضه", "تكنولوجيا المعلومات")

## 5.2 Tools

We utilized the following frameworks and libraries to implement our architecture.

- The **Transformer** library, developed by Hugging Face, is a powerful tool for natural language processing tasks. It enables researchers and developers to incorporate cutting-edge transformer models into their projects. Within the library, the AraBERT model is a pre-trained language model designed explicitly for Arabic text, providing substantial benefits in understanding and generating this language. At aubmindlab, the AraBERT model is applied to various applications, including sentiment analysis, text classification, and named entity recognition, making it a valuable resource for Arabic language processing. The model's architecture allows it to understand the complexities of Arabic syntax and semantics, leading to more accurate and context-aware predictions. In this study, we aim to utilize AraBERT with the Transformer library.
- **UMAP (Uniform Manifold Approximation and Projection) [23]** effectively visualizes high-dimensional embedding data, helping researchers understand complex relationships within their studies' datasets. By reducing the dimensions of embedding vectors, UMAP makes it easy to create clear 2D or 3D visualizations that reflect the underlying structure of the data. UMAP visualization can discover interpretable clusters between data points, simplifying exploratory data analysis and deepening understanding of relationships between different embeddings. **Figures 2, 3, 4, and 5** show the visualization of the learned vectors after reducing them to two dimensions using the UMAP dimensionality reduction technique. A specific color and shape represent each vertical search result. Furthermore, vectors capture valuable semantic information about documents and their interactions.

➢ **TensorFlow [24]** is a versatile and powerful library for building and training machine learning models; thus, it is very suitable for applying stacked autoencoders to embedded data. TensorFlow's excellent support for GPU acceleration also ensures that training stacked autoencoders on large embedding datasets is efficient and scalable. In this paper, we implemented our proposed stacked autoencoder model using TensorFlow.

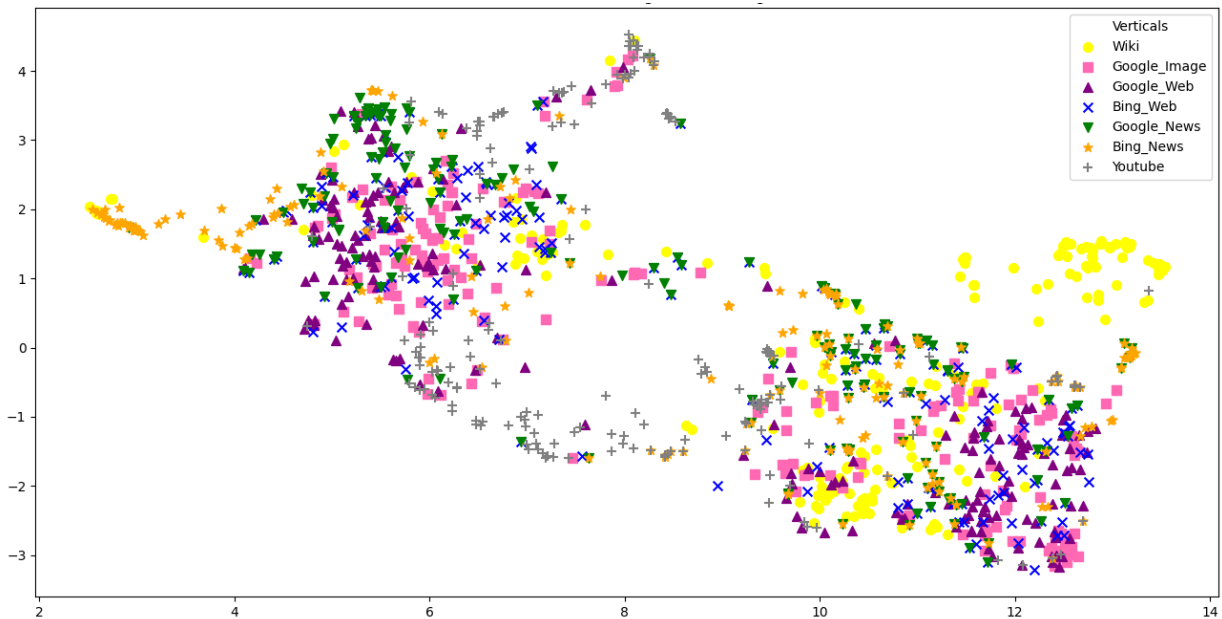

*Fig. 2: Visualizing vector embedding results using UMAP for Q1 ("التعليم", "الرياضه")*

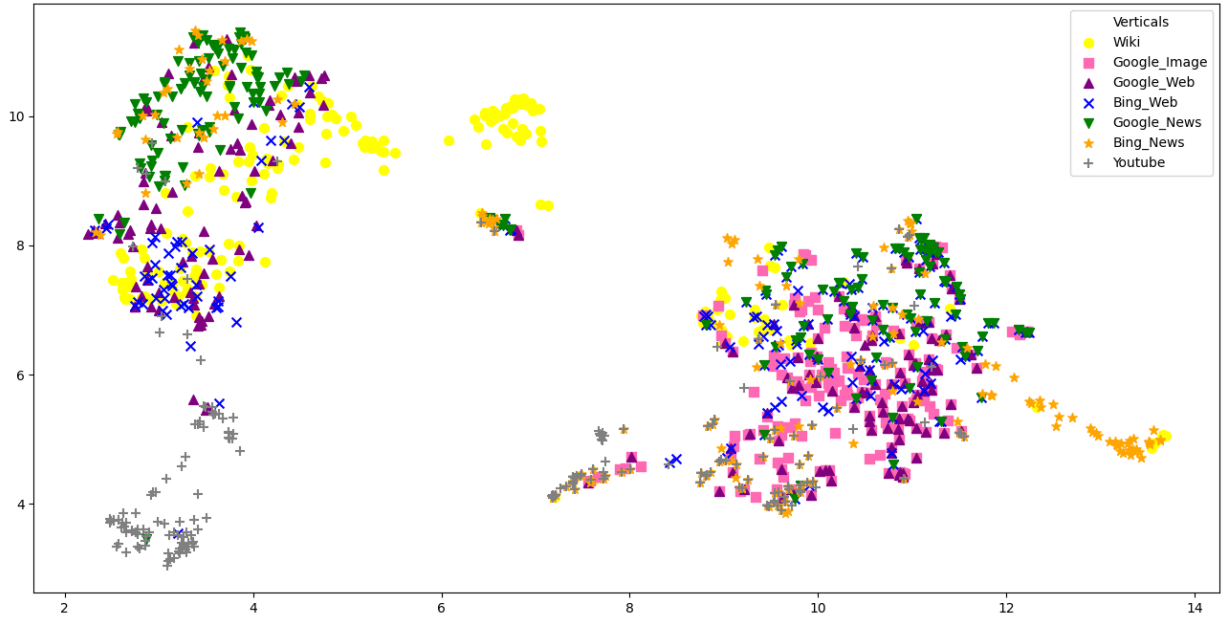

*Fig. 3 Visualizing vector embedding results using UMAP for Q2 ("الرياضه" , "تكنولوجيا المعلومات")*

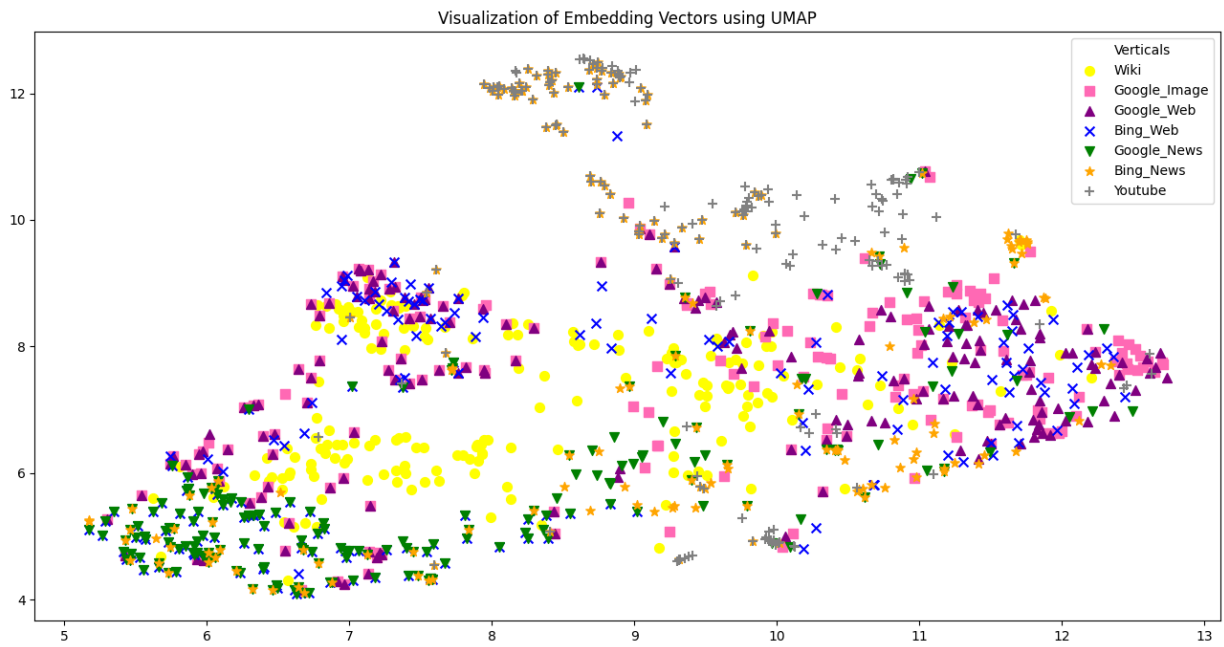

*Fig. 4: Visualizing vector embedding results using UMAP for Q3 ("تكنولوجيا التعليم" , "التعليم")*

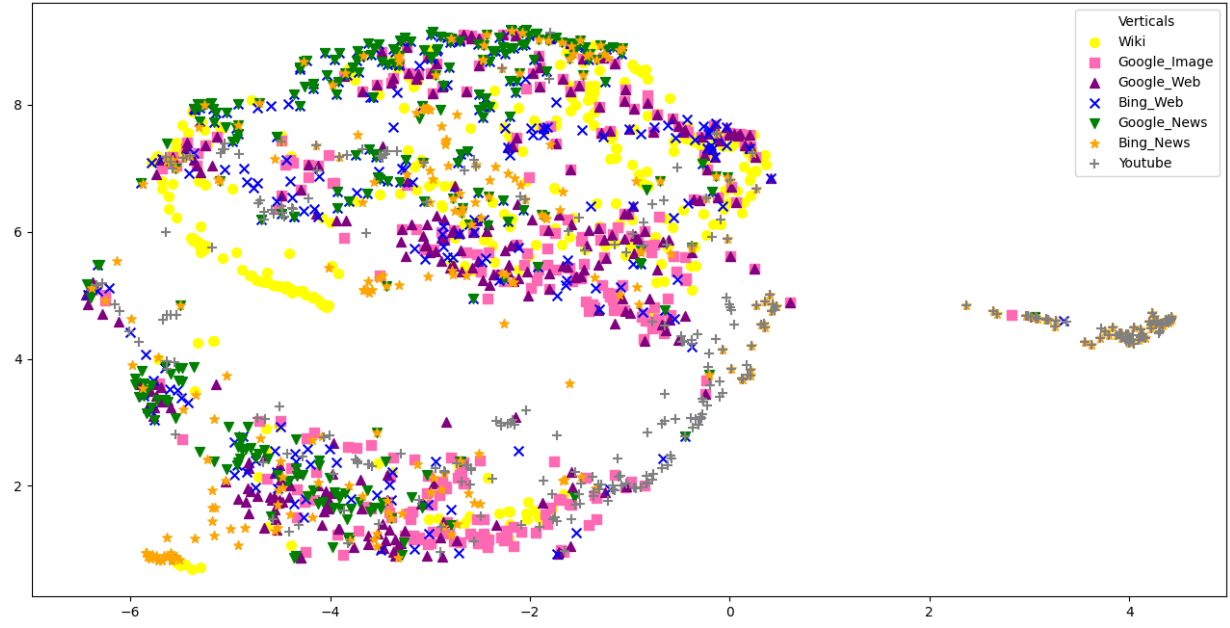

*Fig. 5 Visualizing vector embedding results using UMAP for Q4("الرياضه" , "التعليم", "تكنولوجيا المعلومات")*

The entire network is designed to reconstruct its inputs using a stacked autoencoder architecture, encouraging the hidden layers to learn effective data representations. To support this process, we utilized several key parameters outlined in Table 1.

*Table 1: Parameters for Stacked Autoencoder*

| Parameter | Value |
|---|---|
| Loss function | Mean Squared loss |
| Layer dimension | 512, 256, 128, 64, 32 |
| Learning rate | 0.001 |
| Batch size | 256, 128 |
| Hidden layer | ReLU |
| Output layer | Sigmoid |
| Epochs | [20, 30, 40] |

A plot of the loss and accuracy functions is essential in ascertaining an autoencoder's performance during training. The loss plot shows the difference between predicted and actual outputs, marking model convergence; a downward trend curve suggests error minimization success. The accuracy plot communicates the efficiency of the model in making correct reconstructions. Assessing these metrics helps determine training time, potential overfitting, and overall model performance, so this visualization step is essential to understanding training dynamics and optimization, as in Figs. 6, 7, 8, and 9. We used the **Matplotlib** Python library to create static, animated, and interactive visualizations. Matplotlib can make simple things easy and complex things possible.

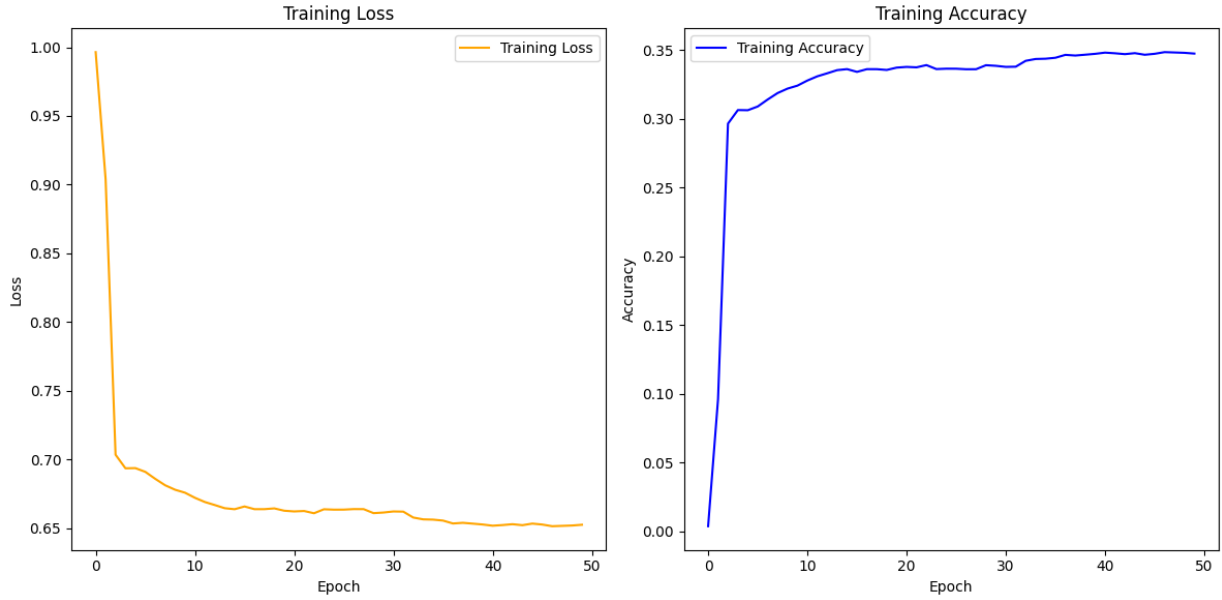

**Fig. 6: Training Accuracy and Error for Q1 per epoch ("الرياضه", "التعليم")**

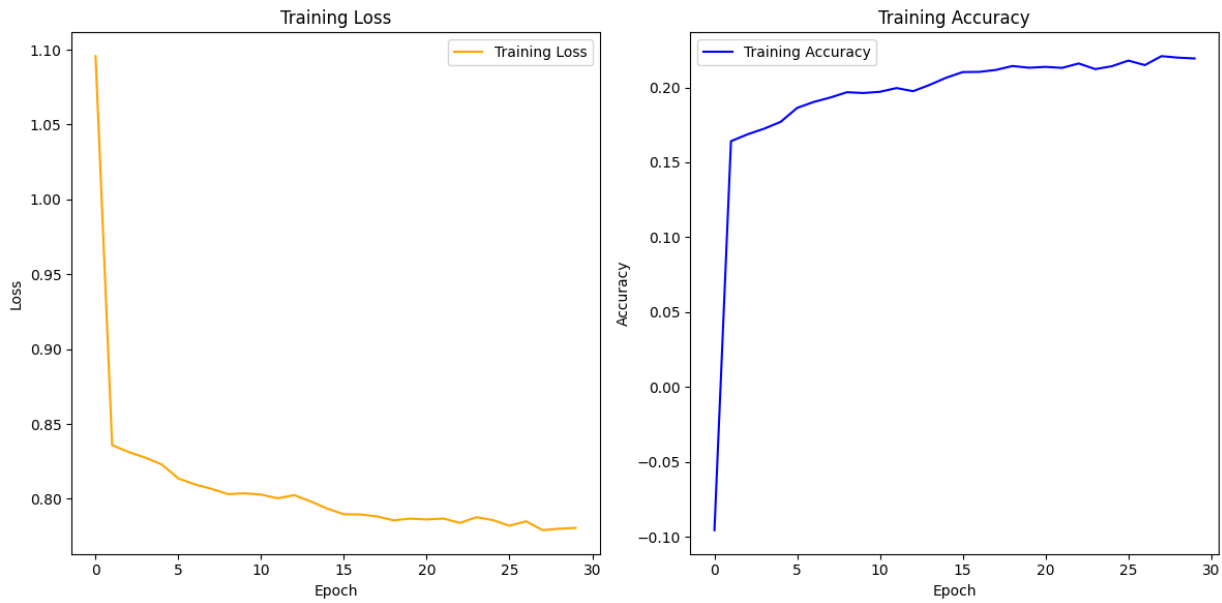

**Fig. 7: Training Accuracy and Error for Q2 per epoch ("تكنولوجيا المعلومات", "الرياضه")**

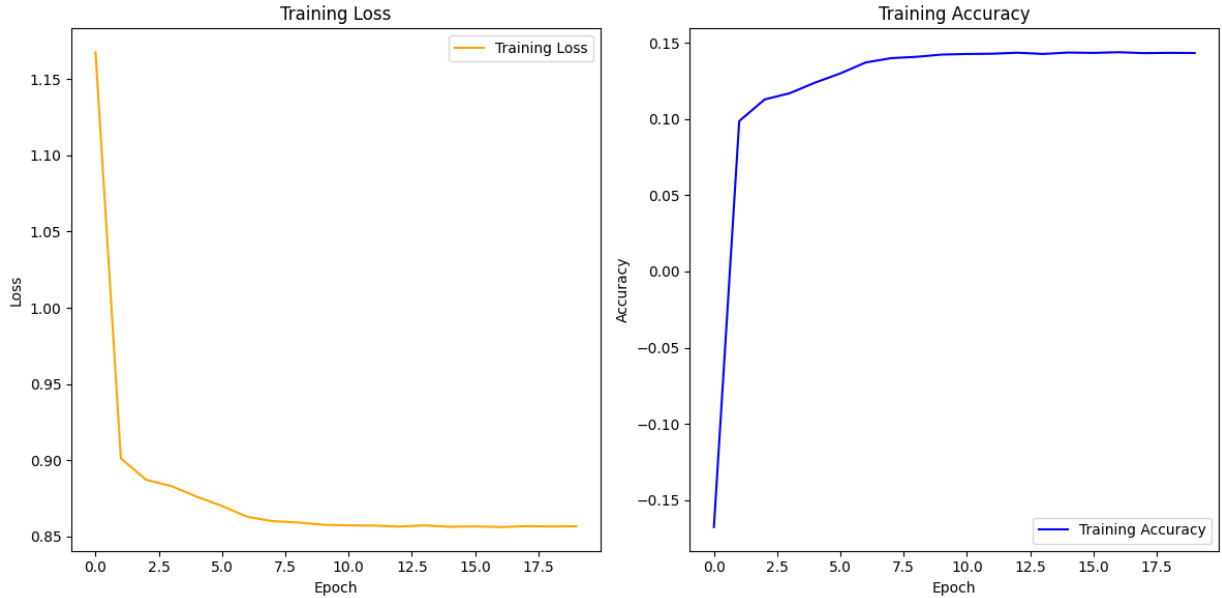

*Fig. 8: Training Accuracy and Error for Q3 per epoch ("التعليم" , "تكنولوجيا المعلومات")*

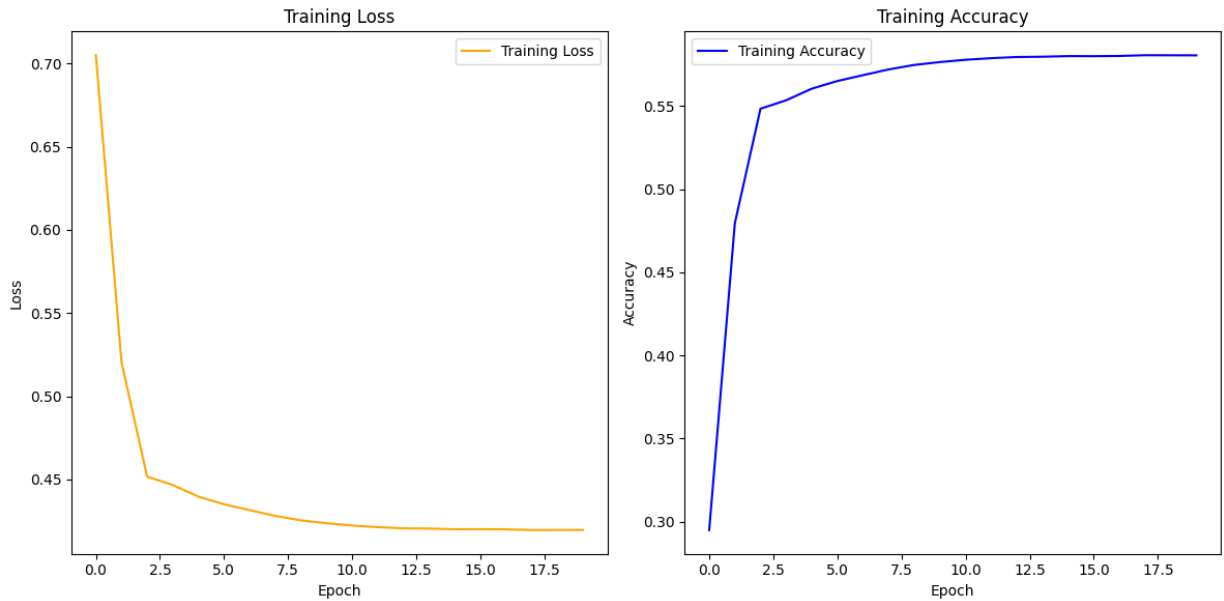

*Fig. 9: Training Accuracy and Error for Q4 per epoch ("تكنولوجيا المعلومات", "الرياضه", "التعليم")*

Finally, we clustered the features retrieved from the previous step (stacked autoencoder) using K-Means, which allows us to group similar data points based on the learned representations. K-Means can effectively identify patterns and clusters in the data using the high-quality embeddings generated by the autoencoder. When applying K-Means, we specify a certain number of clusters [3, 4], as shown **in Figs. 10, 11, 12, and 13**. This allows the algorithm to partition the embedding space into different groups. This process helps reveal the underlying structure and facilitates data segmentation and anomaly detection tasks.

Table 2 shows the number of clusters for the K-means algorithm for four queries. Where the number of clusters K is computed according to the number of topics in each query, for example, Q1: ("الرياضه", "التعليم"), the number of clusters K is 3 ("Sport"," Education,"" Else"), Where the label "Else" indicates to it does not belong to each other clusters c1, c2.

*Table 2: Numbers of clusters (K)*

| Query | K | Labels |
|---|---|---|
| Q1("التعليم ", "الرياضه") | 3 | Sport, Education, Else |
| Q2("الرياضه", "تكنولوجيا المعلومات") | 3 | Sport, Information Technology, Else |
| Q3 ("التعليم", "تكنولوجيا المعلومات") | 3 | Information Technology, Education, Else |
| Q4 ("التعليم", "الرياضه ", "تكنولوجيا المعلومات") | 4 | Sport, Education, Information Technology, Else |

The clustering process is enhanced when stacked autoencoders are combined with K-Means because the autoencoder's ability to detect important features improves the quality of the generated clusters. As a result, this method can lead to a more accurate interpretation of data and produce better results in various areas of applications. By identifying the centroids of each cluster, analyzing the similarities between the data points and the cluster centroids, and ranking the data points within each cluster according to cosine similarity, we can identify the top 10 ranked data points in each cluster.

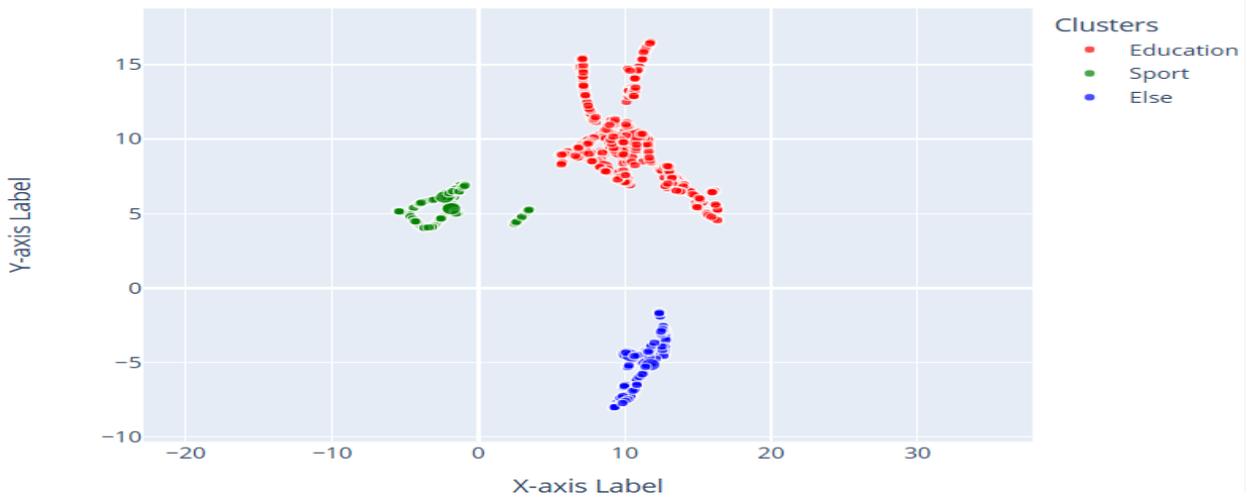

*Fig. 10: The cluster of Q1 ("الرياضه ", "التعليم" )*

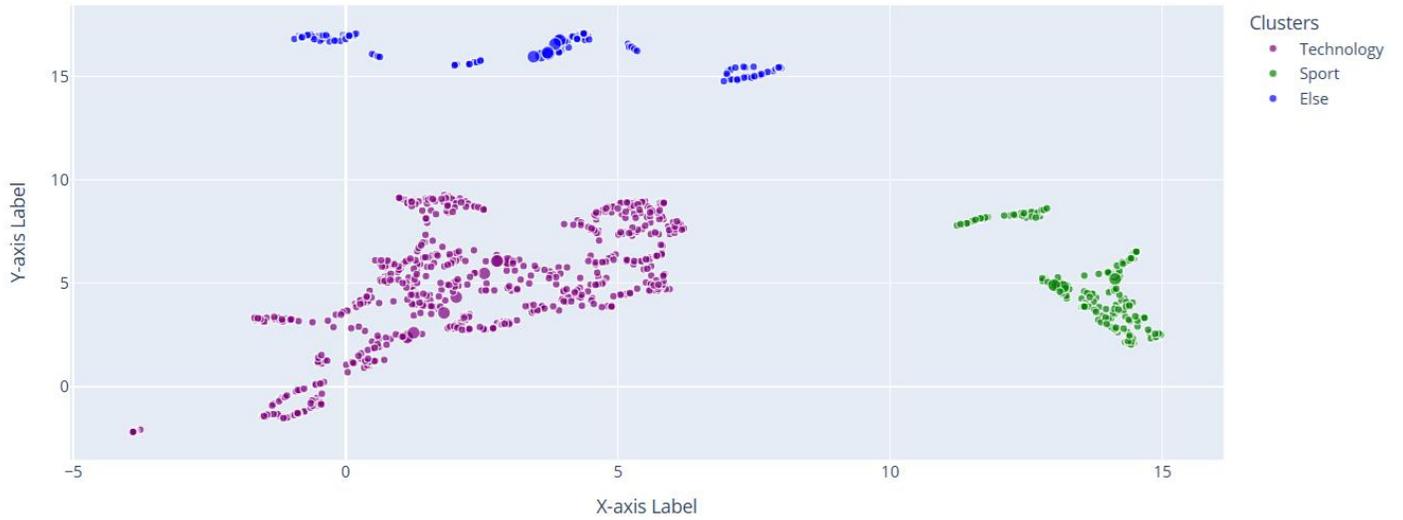

*Fig. 11: The cluster of Q2 ("تكنولوجيا المعلومات" , "الرياضه")*

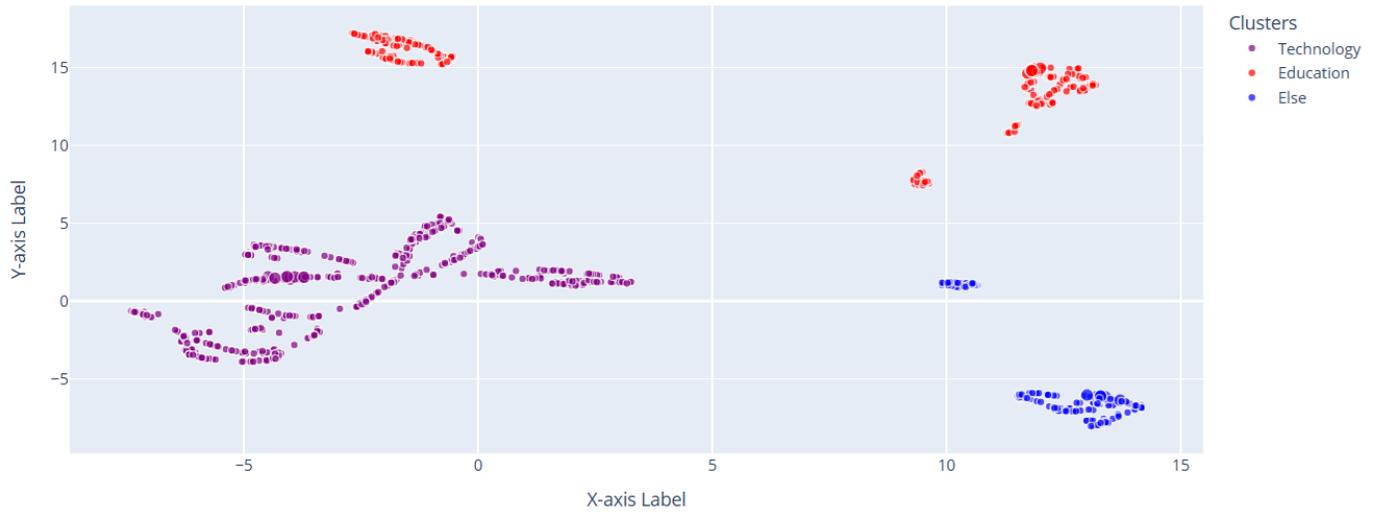

*Fig. 12: The cluster of Q3 ("تكنولوجيا المعلومات" , " التعليم")*

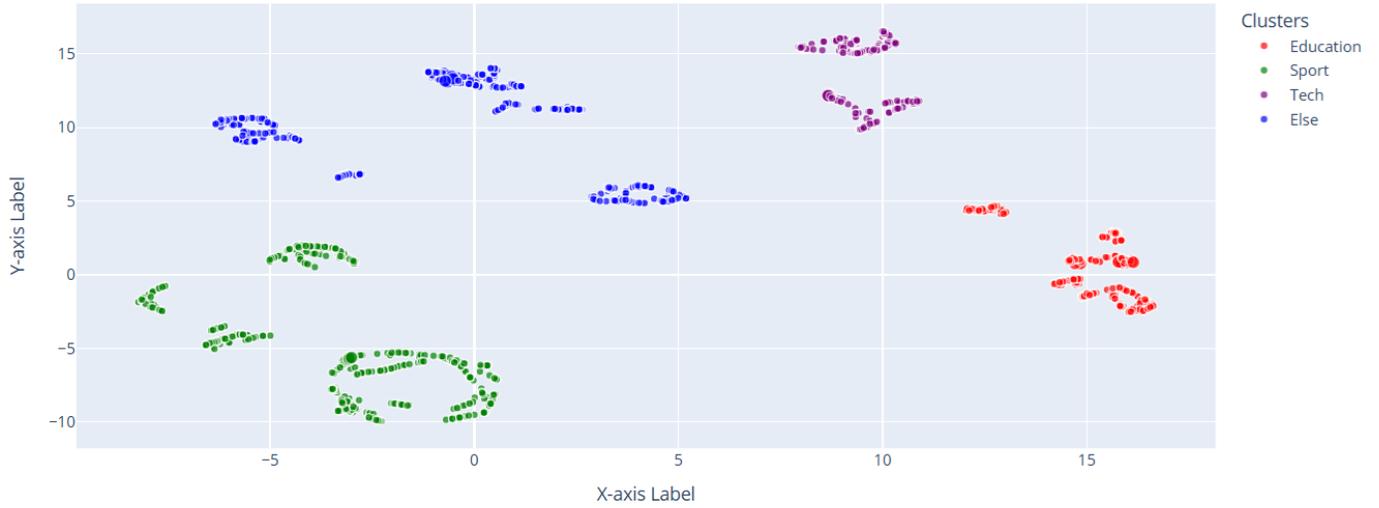

*Fig. 13: The cluster of Q4 ("تكنولوجيا المعلومات", "الرياضه", "التعليم")*

### 5.3 Evaluation metrics

This section outlines the performance metrics employed to assess the proposed method.

1. The Silhouette coefficient measures the separation distance between the generated clusters, ranging from -1 to +1 [25].

Silhouette scores near +1 indicate that adjacent clusters are well separated, while a score of 0 suggests that the clusters are on or near the decision boundary between the two clusters. A silhouette score of -1 implies that the samples were assigned to the wrong cluster.

The formula of the **Silhouette score** is:

$$\textbf{Silhouette Score} = (b - a)/max(a, b)$$

Here, a is the average distance of each point to other points in the same cluster, and b is the average distance to all other clusters. According to Table 3, the K-Means clustering algorithm obtained the average range of silhouette scores for each query.

*Table 3: Silhouette Score for each Query*

| Query | Silhouette score |
|---|---|
| Q1 ("التعليم", "الرياضه") | 0.673 |
| Q2 ("الرياضه", "تكنولوجيا المعلومات") | 0.630 |
| Q3 ("التعليم", "تكنولوجيا المعلومات") | 0.653 |
| Q4 ("تكنولوجيا المعلومات", "الرياضه", "التعليم") | 0.613 |

2. The **Davies-Bolden index** is a validation measure for evaluating clustering models. It is determined by the average similarity of each cluster to its most similar cluster. Here, similarity is defined as the ratio of inter-group to intra-group distances. For every cluster, assess the within-cluster dispersion and between-cluster separation. Lower values imply better clustering.

   The formula of the ***Davies-Bolden index*** is:

   For a dataset X=$X_1, X_2, X_3$, ....The Davies-Bouldin Index for k number of clusters can be calculated as

$$DB = \frac{1}{k}\sum_{i=1}^{k} \max\left(\frac{\Delta(X_i) + \Delta(X_j)}{\delta(X_i, X_j)}\right)$$

Where:

$\Delta(X_k)$ is the intracluster distance within the cluster $X_k$

$\delta(X_i, X_j)$ is the intercluster distance between the clusters $X_i$ and $X_j$

The **Davies-Bolden index** is provided within the sklearn.metrics module of the scikit-learn library. The following syntax should be followed:

**sklearn.metrics.davies_bouldin_score(X, labels)**

According to Table 4, the ***Davies-Bolden index*** was obtained by the K-Means clustering algorithm for each query.

*Table 4: Davies-Bolden Index for each Query*

| Query | Davies-Bolden index |
|---|---|
| Q1 ("الرياضه", "التعليم") | 0.413 |
| Q2 ("تكنولوجيا المعلومات", "الرياضه") | 0.504 |
| Q3 ("تكنولوجيا المعلومات", "التعليم") | 0.608 |
| Q4 ("المعلومات تكنولوجيا", "التعليم", "الرياضه") | 0.573 |

3. The **Dunn index**, introduced by Joseph C. Dunn in 1974, is a measure for evaluating clustering algorithms [26]. It is part of a group of validity indices, including the **Davies-Bolden** Index and the Silhouette score, as it is an internal scoring scheme where the score is based on the clustered data for a given clustering task. The Dunn index measures the ratio of the minimum distance among groups (inter-cluster) to the maximum distance within groups (intra-cluster). A Dunn Index more significant than 1 generally indicates that the clusters are distinct and cohesive.

The formula of the **Dunn index** is:

$$D = \frac{min_{i \neq j}\, d(C_i, C_j)}{max_k\, d(C_k)}$$

Where:

- $D$ is the Dunn Index.
- $C_i$ and $C_j$ represent different clusters.
- $d(C_i, C_j)$ is the distance between clusters $C_i$ and $C_j$.
- $min_{i \neq j}\, d(C_i, C_j)$ is the minimum distance separating any two clusters (inter-cluster distance).
- $d(C_k)$ is the maximum distance between points within a cluster $C_k$ (intra-cluster distance).
- $max_k\, d(C_k)$ is the maximum of the intra-cluster distances across all clusters.

According to Table 5, the Dunn Index was obtained by the K-Means clustering algorithm for each query.

*Table 5: Dunn Index for each Query*

| Query | Dunn Index |
|---|---|
| Q1 ("التعليم ", "الرياضه") | 2.659 |
| Q2 ("الرياضه", "تكنولوجيا المعلومات") | 2.482 |
| Q3 ("التعليم ", "تكنولوجيا المعلومات") | 1.961 |
| Q4 ("الرياضه", "التعليم", "تكنولوجيا المعلومات") | 1.844 |

For Example, According to Tables 3, 4, and 5 for the first query Q1: A **Silhouette score** of 0.673 means that clusters of Q1 are reasonably well-defined, the value of **Davies Bolden index** for the same query Q1 is 0.413, quite good, as it indicates that the clusters are distinctly separated and compact, and the value of **Dunn index** for the same query Q1 is 2.659 suggests that the clusters are separated and that the distances within the cluster are relatively small compared to the distances between the clusters. This is a positive indicator of the quality of the clustering.

# 6 Conclusion

This work integrated an aggregated search with state-of-the-art NLP techniques to improve information retrieval and clustering for Arabic content. We used AraBERT embeddings and stacked autoencoders to transform unstructured data into compact context-aware representations. This allowed us to go beyond the limitations of traditional search engines, such as precision-recall trade-offs, lack of context, and information overload. Our proposed method significantly improved clustering performance using K-Means to discover patterns in heterogeneous vertical search results. Experimental results (silhouette scores) showed that our model can distinguish between clusters and give more insight into Arabic textual data. Future work can focus on refining the clustering algorithms and exploring other embedding techniques to improve the system further. Also, incorporating user feedback into the search process can make the search more personalized, optimized, and user-centric.

**Acknowledgments:** I am deeply grateful to my supervisors for their guidance and support. To my family, thank you for your unwavering love and encouragement.

**Conflict of interest:** Authors declare that neither funding nor conflict of interest exists.

**Data availability:** The datasets generated during and/or analyzed during the current study are available in the [Google Drive] repository,

https://drive.google.com/drive/folders/1cVDzaze-kVukCF66

KuXgrZJN23SB8vbq?usp=sharing.